\documentclass[12pt]{JHEP3}
\usepackage{amsfonts}
\usepackage{bbold}

\usepackage{amsmath}
\usepackage{slashed}
\usepackage{graphicx}

\title{An example of localized D-branes solution on PP-wave backgrounds}
\preprint{\hepth{0301253}\\SISSA 8/2003/EP}
\author{L. F. Alday$^{1,2}$ ,M. Cirafici$^{1}$
\\
$^1$International School for Advanced Studies, Trieste, Italy. \\
$^2$The Abdus Salam ICTP, Trieste, Italy.\\
Email: \, \email{alday@sissa.it},\, \email{cirafici@sissa.it}}

\abstract{In this note we provide an explicit example of type IIB supersymme-\ tric D3-branes solution on a pp-wave like background, consisting in the product of an eight-dimensional pp-wave times a two-dimensional flat space. An interesting property of our solution is the fully localization of the D3-branes (i.e. the solution depends on all the transverse coordinates). Then  we show the generalization to other Dp-branes and to the D1/D5 system.}

\begin{document}

%%%%%%%%%%%%%%%%%%%%%%%%%%%%%%%%%%%

\section{Introduction}

%%%%%%%%%%%%%%%%%%%%%%%%%%%%%%%%%%%

In the last years a duality, relating string theory living on specific
backgrounds (of the form AdS${}_{d}\times M{}_{10-d}$, involving $d$%
-dimensional Anti de Sitter space) and quantum field theories living in the $%
d-1$ dimensional boundary of the AdS${}_{d}$, has been discovered. This
correspondence, and its generalizations, are commonly referred to as the
AdS/CFT correspondence or Maldacena conjecture \cite{malda1}\cite{witten1}\cite{polyakov1}.

This conjecture has very important implications on a wide range of issues in
theoretical physics. On the one hand, it is expected to play a fundamental
role in the eventual non-perturbative formulation of string theory, while on
the other hand it has given us a tool to describe QFTs in the
non-perturbative regime. There is currently a vast literature on this
subject, exploring various directions, for a review see \cite{maldareview}
and references therein.

Very recently a related duality has been discovered, by considering the
AdS/CFT duality in a specific (Penrose) limit (blowing up the geometry in
the neighborhood of a null geodesic \cite{Penrose}), between strings
propagating in a plane parallel (pp) wave background and a special sector of
supersymmetric Yang Mills theories \cite{BMN}. This background has been
proved to be a maximally supersymmetric solution of type IIB string theory 
\cite{Blau1} and also exactly solvable \cite{Metsaev1}.

A fundamental role in the AdS/CFT correspondence is played by D-branes,
since they are, supposedly, dual to non perturbative objects on the gauge
theory side. Hence it is interesting to address the question of their
presence, together with open strings, in this kind of backgrounds. This
problem has already been considered by many authors, in particular see \cite
{Meessen}\cite{parvizi}\cite{skenderis}\cite{morales}\cite{leo} for the study of
D-branes and \cite{gavanarain}\cite{chu}\cite{bala} for the open string
sector of the theory.

In this note we find a supersymmetric solution of the Type IIB string
theory equations corresponding to a fully localized D3-brane in presence of
a pp-wave (where mass is given only to the six transverse scalars), by explicitly solving the equations of motion and the
supersymmetry conditions. We also show how our results can be generalized to
other Dp-branes and to the D1/D5-brane system.

Throughout this note we follow the approach of \cite{Meessen}, from which we
borrow some results; however we use a different ansatz, inspired by a
proposal made by Maldacena and Maoz, in \cite{mm}.

This note is organized as follows: in the next section by analogy with the
general problem of putting D-branes on flat space-time we state our ansatz
for D3--branes on a pp-wave. In section three we show that our solution is supersymmetric and in section four we explicitly write and
solve the equations of motion. In section five we show how our results can
be generalized for other kind of Dp-branes, namely $p=1,5$ and $7$ and in section six we extend our analysis to the D1/D5-brane system. We
finish with some brief concluding remarks.

%%%%%%%%%%%%%%%%%%%%%%%%%%%%%%%%%%%

\section{Putting D-branes on the pp-wave}

%%%%%%%%%%%%%%%%%%%%%%%%%%%%%%%%%%%

Let us briefly review the case of a Dp-brane in flat space-time. Due to the presence of the Dp-brane, the ten dimensional
flat metric:

\begin{equation}
ds^{2}=\eta _{MN}dx^{M}dx^{N}
\end{equation}
is modified as (in the string frame):

\begin{equation}
ds_{s}^{2}=H_{p}^{-1/2}\eta _{\mu \nu }dx^{\mu }dx^{\nu
}-H_{p}^{1/2}dx^{i}dx^{i}
\end{equation}
where $\mu =0,\dots ,p$ runs over the coordinates on the Dp-brane and $%
i=p+1,\dots ,9$ on the transverse coordinates and $H_{p}$ is an harmonic
function of the transverse coordinates. In order to be consistent with the
equations of motion we are obliged to turn on a $(p+2)$ field strength
(whose potential couples to the world volume of the Dp-brane):

\begin{equation}
F_{[p+2]}=g_{s}^{-1}dx^{0}\wedge \dots \wedge dx^{p}\wedge d(H_{p}^{-1})
\end{equation}
(if $F$ is a 5 form we should add also its dual). Further, the
standard relation between the dilaton and the string coupling constant ( $%
e^{2\phi }={g_{s}}^{2}$) is modified as:

\begin{equation}
e^{2\phi }=g_{s}^{2}H_{p}^{\frac{3-p}{2}}  \label{dilans}
\end{equation}
Note that for the special case of a D3-brane the dilaton is constant, and we
can set it equal to zero. From now on we will set $g_s=1$.

To pass to the so called Einstein frame, we perform the following rescaling:

\begin{equation}
g_{\mu \nu ,s}=e^{\phi /2}g_{\mu \nu ,e}
\end{equation}
For a generic $p$, the metric in this frame takes the form:

\begin{equation}
ds_{e}^{2}=H_{p}^{\frac{p-7}{8}}\eta _{\mu \nu}dx^{\mu }dx^{\nu }-H_{p}^{%
\frac{p+1}{8}}dx^{i}dx^{i}
\end{equation}
Note that for $p=3$ the metric is the same in both frames.

The standard pp-wave background reads (written in Brinkman coordinates):

\begin{subequations}
\begin{eqnarray}
ds^{2} &=&2dudv+2S(u,x^{i})dudu-dx^{i}dx^{i} \\
F_{[5]} &=&du\wedge \varphi _{[4]}(x^{i})
\end{eqnarray}
\end{subequations}

Where the light--cone coordinates have been introduced as $u=\frac{%
x^{0}+x^{9}}{2}$ and $v=\frac{x^{0}-x^{9}}{2}$, and $i=1,\dots ,8$. Here $%
\varphi _{[4]}(x^{i})$ is a four form such that the five form is self dual.
This kind of backgrounds was studied previously for instance in \cite{mm} 
\cite{bs}, where the problem of supersymmetry have been addressed.

Many people have studied the problem of putting D-branes in backgrounds of
this kind; however, we will focus on D3-branes and consider the following pp-wave
background instead of the usual one, and try to mimic the procedure
previously explained for the flat case:

\begin{subequations}
\label{backnobrane}
\begin{eqnarray}
ds^{2}&=&\left( 2du\left( dv+S(\vec{y}) du\right) -d\vec{x}%
^{\,2}\right) -d\vec{y}^{\,2} \\
F_{[5]B}&=&\frac{1}{\sqrt2}(\varphi _{\lbrack 3]})\wedge dz^{4}\wedge du+c.c.
 \, \\
\varphi _{\lbrack 3]}&=&W_{1}d{\bar{z}}^{1}\wedge dz^{2}\wedge
dz^{3}+W_{2}dz^{1}\wedge {d\bar{z}}^{2}\wedge dz^{3}+W_{3}dz^{1}\wedge
dz^{2}\wedge {d\bar{z}}^{3}
\end{eqnarray}
\end{subequations}
where now $\vec{x}$ ($z^{4}$ in complex notation) denotes points on $\mathbb{%
R}^{2}$ and $\vec{y}$ \ ($z^{1,2,3}$ in complex notation) points on $\mathbb{%
R}^{6}$. Note that $S$ is now function only of the $\vec{y}$ coordinates.

Inspired by the previous discussion it is natural to propose the following
ansatz for a D3-brane on the background (\ref{backnobrane}):

\begin{subequations}
\label{metric}
\begin{eqnarray}
ds^{2}&=&H\left( \vec{y}\right) ^{-\frac{1}{2}}\left( 2du\left( dv+S\left(
\vec{y}\right) du\right) -d\vec{x}^{\,2}\right) -H\left( \vec{y}\right) ^{\frac{1}{2}}d%
\vec{y}^{\,2}  \\
F_{[5]D}&=&\frac{1}{\sqrt2}  du\wedge dv\wedge dx^{1}\wedge dx^{2}\wedge dH^{-1}\ 
+dual  \\
F_{[5]B}& =&\frac{1}{\sqrt2}(\varphi _{\lbrack 3]})\wedge dz^{4}\wedge du+c.c. \\
\varphi _{\lbrack 3]}&=&W_{1}d{\bar{z}}^{1}\wedge dz^{2}\wedge
dz^{3}+W_{2}dz^{1}\wedge {d\bar{z}}^{2}\wedge dz^{3}+W_{3}dz^{1}\wedge
dz^{2}\wedge {d\bar{z}}^{3}
\end{eqnarray}
\end{subequations}

Consistently with (\ref{dilans}) we are assuming constant dilaton. Note that
we added the dual of the D-brane 5 form, as before. The $W^{\,\prime }$s have to be chosen in such a way
that the five-form is closed. Here we are placing the D3-branes on the
coordinates $u,v$ and $\vec{x}$ (or, in complex notation, in the $4^{th}$
complex plane) corresponding to a \textit{longitudinal} D-brane \cite
{Meessen}. Of course there are other possibilities, namely \textit{\
transverse} (that is with the $u$ coordinate belonging to the brane world volume but not $v$) or \textit{instantonic} (neither $u$ nor $v$ belonging to the brane world volume ) D-branes \cite{skenderis} \cite{bgg}.
Since in our case with this orientation the worldsheet scalars coming from
the pp--wave are transverse to the D3-brane this choice seems to be the
natural one.

The next sections are devoted to the study of this ansatz, first looking at
the conditions for existence of supersymmetry generators and then solving
the equations of motion.

%%%%%%%%%%%%%%%%%%%%%%%%%%%%%%%%%%

\section{Supersymmetry conditions}

%%%%%%%%%%%%%%%%%%%%%%%%%%%%%%%%%%

In this section we will analyze in some detail the supersymmetry conditions
for the background (\ref{metric}). The supersymmetries are obtained by
equating the variations of the gravitino and the dilatino to zero and
looking for non-trivial spinors $\left( \epsilon \right) $ satisfying these
restrictions. The generic supersymmetry variations for the dilatino and
gravitino in the doubled formulation of supergravity (both electric and magnetic RR fields, except for $C_{[0]}$, are used) are:

{\footnotesize
\begin{subequations}
\label{susy}
\begin{eqnarray}
\delta \Psi _{\mu } &=&D_{\mu }\epsilon +\frac{1}{16}e^{\phi }\left(
2\slashed\partial C_{\left[ 0\right] }\left( i\sigma ^{2}\right) +\frac{1}{3!}%
\slashed F_{[3]}\left( \sigma ^{1}\right) +\frac{1}{5!}\slashed %
F_{[5]}\left( i\sigma ^{2}\right) +\frac{1}{7!}\slashed F_{[7]}\left( \sigma
^{1}\right) \right) \gamma _{\mu }\epsilon  \\
\label{dilatinovar}
\delta \chi  &=&\slashed\partial \phi \epsilon +\frac{1}{4}e^{\phi }\left(
-4\slashed\partial C_{\left[ 0\right] }\left( i\sigma ^{2}\right) -\frac{1}{3!}%
\slashed F_{[3]}\left( \sigma ^{1}\right) +\frac{1}{7!}\slashed %
F_{[7]}\left( \sigma ^{1}\right) \right) \epsilon 
\end{eqnarray}
\end{subequations}
}

With $F$ we denote the RR forms of type IIB string theory, $C_{[0]}$ is the
axion and $\phi $ the dilaton. We define $\slashed F_{[n]}=F_{\alpha
_{1}\dots \alpha _{n}}\Gamma ^{\alpha _{1}\dots \alpha _{n}}$. The covariant
derivative is defined as:

\begin{equation}
D_{\mu }=\partial _{\mu }-\frac{1}{4}\omega _{\mu \underline{ab}}\Gamma ^{%
\underline{ab}}
\end{equation}
and the spin connection $\omega$ as function of the vielbein given by: 

\begin{equation}
e_{\nu }{}^{\underline{m}}e_{\rho }{}^{\underline{n}}\omega _{\mu \underline{mn}}=\frac{1%
}{2}[e_{\rho \underline{p}}(\partial _{\mu }e_{\nu }{}^{\underline{p}%
}-\partial _{\nu }e_{\mu }{}^{\underline{p}})-e_{\mu \underline{p}}(\partial
_{\nu }e_{\rho }{}^{\underline{p}}-\partial _{\rho }e_{\nu }{}^{\underline{p}%
})+e_{\nu {\underline{p}}}(\partial _{\rho }e_{\mu }{}^{\underline{p}%
}-\partial _{\mu }e_{\rho }{}^{\underline{p}})]
\end{equation}

With $\mu ,\nu ,etc$ we indicate space-time coordinates and with $\underline{a},\underline{b},etc$ coordinates on the tangent space. In the following, $a$ will refer to coordinates transverse to the D-brane whereas $i$ to
coordinates longitudinal to the D-brane. The components of the vielbein are
chosen as: $\left( e^{\underline{m}}={e_{\mu }}^{\underline{m}}dx^{\mu
}\right) $

\begin{equation}
\begin{array}{ll}
e_{\underline{v}}=e^{\underline{u}}=H^{-{\frac{1}{4}}}du\,, & e_{\underline{u%
}}=e^{\underline{v}}=H^{-{\frac{1}{4}}}(dv+Sdu)\,,
\\[3mm]
-e_{\underline{i}}=e^{\underline{i}}=H^{-{\frac{1}{4}}}dx^{i}, & -e_{%
\underline{a}}=e^{\underline{a}}=H^{\frac{1}{4}}dy^{a}
\end{array}
\end{equation}
and of the inverse vielbeins as: $\left( \theta _{\underline{m}}={e_{\underline{%
m}}}^{\mu }\partial _{\mu }\right) $
 
\begin{equation}
\begin{array}{ll}
\theta _{\underline{u}}=H^{\frac{1}{4}}(\partial _{u}-S\partial _{v})\,, & 
\theta _{\underline{v}}=H^{\frac{1}{4}}\partial _{v}\,, \\[3mm] 
\theta _{\underline{i}}=H^{\frac{1}{4}}\partial _{i}, & \theta _{\underline{a%
}}=H^{-{\frac{1}{4}}}\partial _{a}
\end{array}
\end{equation}
and we use flat light cone metric $\eta ^{\underline{uv}}=1$, $\eta ^{%
\underline{ij}}=-\delta _{ij}$ and $\eta ^{\underline{ab}}=-\delta _{ab}$.
The non-zero components of the spin connection are: 

\begin{equation}
\begin{array}{ll}
\omega _{%
\underline{u}\underline{u}\underline{a}}=+H^{-{\frac{1}{4}}}\partial _{a}S\,,\,\,\,
\omega _{\underline{uva}}=-{\frac{1}{4}}H^{-{\frac{5}{4}}}\partial _{a}H\,\,,\\[3mm]
 \omega _{\underline{vua}}=-\frac{1}{4}H^{-{\frac{5}{4}}}\partial _{a}H\,,\,\,\,
\omega _{\underline{iaj}}=-\frac{1}{4}H^{-{\frac{5}{4}}}\partial _{a}H\delta
_{ij}\,, \\[3mm]  \omega _{\underline{abc}}=-{\frac{1}{2}}H^{-{\frac{5}{4}}}\eta
_{a[b}\partial _{c]}H
\end{array}
\end{equation}

For the background (\ref{metric}) the dilatino variation is automatically zero, whereas the
gravitino variation becomes:

\begin{equation}
\delta \Psi _{\mu }=D_{\mu }\epsilon +\frac{1}{16}\left( \frac{1}{5!}%
\slashed F_{[5]}\left( i\sigma ^{2}\right) \right) \gamma _{\mu }\epsilon
\end{equation}

Equating this variation to zero we obtain one equation for every direction,
the solutions of whose are the killing spinors. The resulting
equations are:

\begin{subequations}
\label{susycond}
\begin{eqnarray}
\label{adir}
\partial _{a}\epsilon +\frac{1}{8}H^{-1}(\partial _{a}H)\epsilon
+(H^{-1/4}\mathcal{W}_{[5]}(i\sigma ^{2}))H^{1/4}\gamma _{\underline{a}}\epsilon & =0\\
\partial _{v}\epsilon & =0 \\
\partial _{i}\epsilon +(H^{-1/4}\mathcal{W}_{[5]}(i\sigma ^{2}))H^{-1/4}\gamma _{%
\underline{i}}\epsilon & =0 \\
\partial _{u}\epsilon +(H^{-1/4}\mathcal{W}_{[5]}(i\sigma ^{2}))H^{-1/4}\gamma ^{%
\underline{v}}\epsilon -\frac{1}{2}H^{-1/2}\partial _{a}S\Gamma ^{\underline{%
ua}}\epsilon & =0
\end{eqnarray}
\end{subequations}

With $\mathcal{W}_{[5]}$ we denote the
contribution coming from the background 5-form where the $H$ dependence has
been explicitly shown. In order to obtain these equations we also assumed the standard chirality condition in the D3-brane world volume:

\begin{equation}
\Gamma_{wv}(i\sigma^2)\epsilon=\epsilon
\end{equation}
where $\sigma^2$ acts on $\epsilon$ as a doublet of 16 components complex spinor.
By rescaling the spinor with a factor of $H^{-1/8}$ the equation (\ref{adir}%
) becomes \footnote{%
This kind of rescaling is standard in the flat space D-brane solution.}:

\begin{equation}
\partial _{a}\epsilon +(H^{-1/4}\mathcal{W}_{[5]}(i\sigma ^{2}))H^{1/4}\gamma _{%
\underline{a}}\epsilon =0\,
\end{equation}
whereas, since $H$ only depends on $\vec{y}$, the other equations remain untouched.

A solution can be easily found by considering constant spinors. Let us
denote them by $(\pm \frac{1}{2},\pm \frac{1}{2},\pm \frac{1}{2},\pm \frac{1%
}{2},\pm \frac{1}{2})$, where every sign corresponds to the chirality with
respect to the corresponding complex plane (or to the $u$--$v$ plane). We choose conventions such that $\gamma^{\underline{u}}(+)=0$, $\gamma^{\underline{v}}(-)=0$, $\Gamma^{i}(+)=0$ and $\Gamma^{\overline{i}}(-)=0$, where $\Gamma^i={%
\Gamma^{\bar i}}^*=\gamma^x +i \gamma^y$ are the complex $\gamma$ matrices for
the complex plane $i$. 

The condition of negative space-time chirality implies an odd number of minus signs \footnote{In our conventions $\Gamma^{11}=\Gamma^{uv12345678}=\Gamma^{uv1\overline{1}2\overline{2}3\overline{3}4\overline{4}}$}. 

Remember that equations (\ref{susycond}) were obtained imposing a definite chirality in the
world volume of the D-brane. This means an even number of minuses in the two
planes on which the D3-brane lies (in our case, the first and the last
planes) in the case of positive chirality and an odd number in the case of negative.

A trivial solution can be found by noticing that $\gamma^{\underline u}$
annihilates the spinors with a plus in the first position, and $\mathcal{W}_{[5]}$ will annihilate the
spinors of the form $(\pm \frac{1}{2}, + \frac{1}{2}, + \frac{1}{2},+ \frac{1%
}{2}, \pm \frac{1}{2})$ and $(\pm \frac{1}{2}, - \frac{1}{2}, - \frac{1}{2}%
,- \frac{1}{2}, \pm \frac{1}{2})$, since a definite sign (plus or minus)
will be killed either by $\Gamma^i$ or by $\Gamma^{\bar i}$. So, we see that the spinor $(+ \frac{1}{2}, - \frac{1%
}{2}, - \frac{1}{2},- \frac{1}{2}, + \frac{1}{2})$ is a solution for positive world-volume chirality, and  $(+ \frac{1}{2}, + \frac{1%
}{2}, + \frac{1}{2},+ \frac{1}{2}, - \frac{1}{2})$ for negative world volume chirality.

However this is not the only solution. We can find another solution
supposing that $\partial _{i}\epsilon =0$ and $\partial _{u}\epsilon =0$ so
that equations (\ref{susycond})\ become:

\begin{subequations}
\label{eqmm}
\begin{eqnarray}
\partial _{a}\epsilon +\mathcal{W}_{[5]}(i\sigma ^{2})\gamma _{\underline{a}}\epsilon &
=0 \\
\partial_{v} \epsilon &= 0\\
\mathcal{W}_{[5]}(i\sigma ^{2})\gamma _{\underline{i}}\epsilon & =0\, \\
\mathcal{W}_{[5]}(i\sigma ^{2})\gamma ^{\underline{v}}\epsilon -\frac{1}{2}\partial
_{a}S\Gamma ^{\underline{ua}}\epsilon & =0\,
\end{eqnarray}
\end{subequations}

We stress the remarkable fact that the function $H$ has disappeared from (\ref
{susycond}); because of this we obtain exactly the \cite{mm} equations \footnote{Equations (\ref{eqmm}) should be compared with (2.5) of \cite{mm} taking into account the difference of conventions, for instance, there $\epsilon$ is a 16 components complex spinor.}, with the
further restriction of definite chirality on the D-brane worldvolume 
\footnote{%
As it can be checked this condition is consistent, since the chirality
matrix leaves equations (\ref{eqmm}) invariant.} and the condition that the
spinor is function only of the transverse coordinates to the D-brane.

In particular our background will be a special case of the background (2.9) of \cite{mm}

\begin{equation}
\begin{array}{ll}
ds^2=-2dudv-32(|\partial_k \mathtt{W}|^2+|\phi_{j \bar k}z^j|^2)(du)^2+dz^id\bar{z}^i\\[3mm]
\phi_{mn}=\partial_m \partial_n \mathtt{W},\,\,\,\,\,\,\phi_{\overline{mn}}=\partial_{\bar m} \partial_{\bar n} \overline{\mathtt{W}},\,\,\,\,\,\, \phi_{l \bar m}=constants
\end{array}
\end{equation}

where $\phi_{mn}$ and $\phi_{\overline{mn}}$ are defined in terms of $\varphi_{[4]}$ as:

\begin{equation}
\begin{array}{ll}
\phi_{mn}=\frac{1}{3!}(\varphi_{[4]})_{m\overline{ijk}}\epsilon^{\overline{ijkn}}g_{n \bar n}\\
\phi_{\overline{mn}}=\frac{1}{3!}(\varphi_{[4]})_{\overline{m}ijk}\epsilon^{ijkn}g_{n \bar n}
\end{array}
\end{equation}

Since in our background we don't have forms of the kind (2,2) (two holomorphic and two antiholomorphic indices) $\phi_{m \bar n}$ are zero. Here $\mathtt{W}$ is a generic holomorphic function.

In \cite{mm} were considered spinors with $\epsilon_{-}$ \footnote{Note that we use opposite convention for $\epsilon_{+}$ and $\epsilon_{-}$.}:

\begin{equation}
\label{mmspinors}
\epsilon_{-}=\alpha  (- \frac{1}{2}, + \frac{1%
}{2}, + \frac{1}{2},+ \frac{1}{2}, + \frac{1}{2}) + \zeta  (- \frac{1}{2}, - \frac{1%
}{2}, - \frac{1}{2},- \frac{1}{2}, - \frac{1}{2})
\end{equation}
where $\alpha$ and $\zeta$ are complex numbers. Then, for a given $\mathtt{W}$, $\epsilon_{+}$ is solved in function of $\epsilon_{-}$.

With the additional restriction of definite chirality with respect to the world-volume of the D3-brane, we have to choose one and only one of these spinors.

Finally we have (at least) 2 complex spinors (for a definite world-volume chirality), this means 1/8 of supersymmetries.

%%%%%%%%%%%%%%%%%%%%%%%%%%%%%%%%%%%%%%%%%%%%%%%%%%%%%%%%%%%

\section{Equations of motion}

%%%%%%%%%%%%%%%%%%%%%%%%%%%%%%%%%%%%%%%%%%%%%%%%%%%%%%%%%%

For the ansatz (\ref{metric}) the Einstein equations are:

\begin{equation}
R_{\mu \nu }={\frac{1}{2}}\frac{1}{4!}T_{\mu \nu }^{[5]}
\end{equation}
where $T^{[5]}$ is\ the energy momentum tensor for the RR 5-form (the only
different from zero for our background) defined by:

\begin{equation}
T^{[5]}_{\mu\nu}=F_{[5]\mu \alpha 1 ... \alpha 4}F_{[5]\nu}^{\alpha 1 ... \alpha
4}-\frac{1}{10}g_{\mu\nu}(F_{[5]\alpha 1 ... \alpha 5}F_{[5]}^{\alpha 1 ...
\alpha 5})
\end{equation}

For our background we obtain:

\begin{subequations}
\label{tmunu}
\begin{eqnarray}
T_{uu} &=&4!\frac{1}{H}(|W_{1}|^{2}+|W_{2}|^{2}+|W_{3}|^{2})+4!\frac{S%
}{H^{3}}(\partial _{a}H\partial _{a}H) \\
T_{uv} &=&\frac{4!}{2}\left( \frac{\partial _{a}H\partial _{a}H}{H^{3}}\right) \\
T_{vv} &=&0 \\
T_{ij} &=&-\delta _{ij}\frac{4!}{2}\left( \frac{\partial _{a}H\partial _{a}H}{H^{3}}%
\right) \\
T_{ab} &=&-4!\left(\frac{\partial _{a}H\partial _{b}H}{H^{2}}\right)
+\delta _{ab}\frac{4!}{2}\left( \frac{\partial _{c}H\partial _{c}H}{H^{2}}\right)
\end{eqnarray}
\end{subequations}
with the rest of the components equal to zero. Whereas the components of the
Ricci tensor for the metric (\ref{metric}) are:

\begin{subequations}
\label{riccitensor}
\begin{eqnarray}
R_{vv} &=&0 \\
R_{uu} &=&\frac{1}{2H^{3}}(S\left( \partial _{a}H\,\partial _{a}H-H\left( {%
\partial _{a}\partial _{a}}H\right) \right) +2H^{2}({\partial }_{a}\partial
_{a}S))\, \\
R_{uv} &=&\frac{1}{4H^{3}}(\partial _{a}H\,\partial _{a}H-H(\partial
_{a}\partial _{a}H))\, \\
R_{ab} &=&\delta _{ab}\frac{1}{4H^{2}}(\partial _{c}H\,\partial _{c}H-H({%
\partial }_{a}\partial _{a}H))-\frac{1}{2H^{2}}(\partial _{a}H\,\partial
_{b}H)\, \\
R_{ij} &=&-\delta _{ij}\frac{1}{4H^{3}}(\partial _{a}H\,\partial _{a}H-H({%
\partial _{a}}^{2}H))
\end{eqnarray}
\end{subequations}

In (\ref{tmunu}) and (\ref{riccitensor}) the contractions are intended as in euclidean space.

Finally, from the equations of motion we obtain the following conditions \footnote{Similar equations were found, for instance, in \cite{clps}\cite{bonelli}}:

\begin{subequations}
\begin{eqnarray}
\partial_{a}\partial_{a}H& =&0  \label{harm} \\
\label{SWrel}
\partial_{a}\partial_{a}S& =&\frac{1}{2}(|W_{1}|^{2}+|W_{2}|^{2}+|W_{3}|^{2})
\end{eqnarray}
\end{subequations}

The condition (\ref{harm}) just says that $H$ is harmonic in the 6 transverse
directions; this fact\ together with the correct asymptotic behavior for $H$
(far from the D3-brane the space time should look like the standard pp-wave)
gives:

\begin{equation}
H=1+\frac{Q}{y^{4}}
\end{equation}
with $y^{2}=\left( {y^{1}}\right) ^{2}+\dots +\left( {y^{6}}\right) ^{2}$ and Q a non negative real number.

As an important difference with \cite{Meessen} we stress that this solution
is dependent on all the transverse coordinates (and also supersymmetric);
this difference is basically due to our five form.

The remaining equation, (\ref{SWrel}), is easily solved, for instance for a quadratic $S$
(corresponding to mass terms in the pp-wave) and constant $W^{\prime }$s,
giving just a relation among them. This equation admits, however, more
general solutions (of course with the restriction on the $W^{\prime }$s coming from the closure of the five form background).

In the next section we will see how part of these results can
be generalized to other kind of Dp-branes.

%%%%%%%%%%%%%%%%%%%%%%%%%%%%%%%%%%%%%%%%%%%%%%%%%%

\section{Generalization to other Dp-branes}

%%%%%%%%%%%%%%%%%%%%%%%%%%%%%%%%%%%%%%%%%%%%%%%%%%%%%%

In this section we will extend the previous results for other Dp-branes.

First we have to take into account that for other Dp-branes the dilaton is
not a constant anymore. It seems natural to suppose that relation (\ref
{dilans}) holds also for this kind of background. We will also suppose that
the pp-wave background is supported by the following $(p+2)$ form:

\begin{equation}
F_{[p+2]}=du\wedge dv \wedge dx^{1}\wedge \dots \wedge dx^{p-1}\wedge d(H_{p}^{-1})
\end{equation}

Now the dilatino supersymmetry variation (\ref{dilatinovar}) becomes \cite{Meessen}:

\begin{equation}
\delta \chi =\left( \partial _{a}H\gamma ^{\underline{a}}\right)  \left( 1-\Gamma _{wv} \mathcal{P} \right)
\epsilon =0\,  \label{dilatino}
\end{equation}
where $\Gamma _{wv}$ is the chirality matrix on the world-volume of the
Dp-brane we are considering and $\mathcal{P}$, as can be read off from (\ref
{susy}), is a projector given by $\sigma ^{1}$ or $i\sigma ^{2}$ depending on 
$p$ (recall that this projector acts on $\epsilon $ as a doublet of 16 components
complex spinor).

So we see that asking the standard chirality condition with respect to the world-volume of
the Dp-brane :
\begin{equation}
\label{defchirality}
\Gamma _{wv} \mathcal{P}\epsilon=\epsilon
\end{equation}
the equation above is satisfied.

Now we will discuss the issue of the supersymmetric variation of the
gravitino and the equations of motion. The
background to consider for a generic Dp-brane is (in the string frame):

\begin{subequations}
\label{stringf}
\begin{eqnarray}
%ds^{2}&=&H\left( \vec{y}\right) ^{\frac{p-7}{8}}\left( 2du\left( dv+S\left(
%\vec{y}\right) du\right) -d\vec{x}^{\,2}\right) -H\left( \vec{y}\right) ^{\frac{p+1}{8}}d%
%\vec{y}^{\,2}\, \\
ds^{2}&=&H\left( \vec{y}\right) ^{-\frac{1}{2}}\left( 2du\left( dv+S\left(
\vec{y}\right) du\right) -d\vec{x}^{\,2}\right) -H\left( \vec{y}\right) ^{\frac{1}{2}}d%
\vec{y}^{\,2}\, \\
F_{[p+2]D}&=&du\wedge dv\wedge dx^{1}\dots \wedge dx^{p-1}\wedge dH^{-1}
\\
F_{[5]B}&=&\frac{1}{\sqrt 2}(\varphi _{\lbrack 4]})\wedge du+c.c. \\
\varphi _{\lbrack 4]}&=&W_{1}d{\bar{z}}^{1}\wedge dz^{2}\wedge dz^{3}\wedge
dz^{4}+W_{2}dz^{1}\wedge {d\bar{z}}^{2}\wedge dz^{3}\wedge dz^{4}\\
&& + W_{3}dz^{1}\wedge dz^{2}\wedge {d\bar{z}}^{3}\wedge
dz^{4}+W_{4}dz^{1}\wedge dz^{2}\wedge dz^{3}\wedge d{\bar{z}}^{4} \nonumber
\end{eqnarray}
\end{subequations}
(plus the usual relation between the dilaton and $H$) where now $\vec{y}$
denotes points in $\mathbb{R}^{9-p}$ and $\vec{x}$ denotes points in $%
\mathbb{R}^{p-1}$. For a Dp-brane, with $k=\frac{9-p}{2}$ only $W_{1}\dots
W_{k}$ will be different from zero.

For the supersymmetry variation of the gravitino we obtain:

\begin{subequations}
\label{susycondDp}
\begin{eqnarray}
\partial _{a}\epsilon +\frac{1}{8}H^{-1}(\partial _{a}H)\epsilon +(H^{-\frac{1}{4}}\mathcal{W}_{[5]}(i\sigma ^{2}))H^{1/4}\gamma _{\underline{a}}\epsilon &=&0 \\
\partial _{v}\epsilon &=&0 \\
\partial _{i}\epsilon +(H^{-\frac{1}{4}}\mathcal{W}_{[5]}(i\sigma ^{2}))H^{-1/4}\gamma
_{\underline{i}}\epsilon &=&0 \\
\partial _{u}\epsilon +(H^{-\frac{1}{4}}\mathcal{W}_{[5]}(i\sigma ^{2}))H^{-1/4}\gamma
^{\underline{v}}\epsilon -\frac{1}{2}H^{-1/2}\partial _{a}S\Gamma ^{%
\underline{ua}}\epsilon &=&0
\end{eqnarray}
\end{subequations}

These equations are exactly the ones obtained in section 3. There, we solved them for non constant spinors referring to the techniques developed in \cite{mm} and imposing definite chirality in the world volume of the D-brane. In order to redo this analisys for a Dp-brane, we must first check that the condition of definite chirality on the D-brane can be imposed consistently. Unfortunately, this is not the case for D1 and D5 branes, since in these cases equations (\ref{susycondDp}) are not invariant under the action of the chirality matrix $\Gamma_{wv}$: because of this we will only be able to find constant solutions. Let us study these cases separately:

\textbf{D7}: one can easily check that if positive chirality is imposed on the world volume of the brane, the constant spinors $(+ \frac{1}{2}, - \frac{1}{2}, - \frac{1}{2},- \frac{1}{2}, + \frac{1}{2})$, $(+ \frac{1}{2}, - \frac{1}{2}, + \frac{1}{2},- \frac{1}{2}, - \frac{1}{2})$ and $(+ \frac{1}{2}, - \frac{1}{2}, - \frac{1}{2},+ \frac{1}{2}, - \frac{1}{2})$ are solutions. On the other side, for negative chirality, the solutions are  $(+ \frac{1}{2}, + \frac{1}{2}, + \frac{1}{2},+ \frac{1}{2}, - \frac{1}{2})$, $(+ \frac{1}{2}, + \frac{1}{2}, + \frac{1}{2},- \frac{1}{2}, + \frac{1}{2})$ and $(+ \frac{1}{2}, + \frac{1}{2}, - \frac{1}{2},+ \frac{1}{2}, + \frac{1}{2})$. In both cases, we should add (only)  one of the spinors (\ref{mmspinors}). This means that for definite chirality, we have four complex spinors satisfying equations (\ref{susycondDp}), or, in other words, 1/4 of the supersymmetries are preserved.

\textbf{D5}: as stated before, for a D5 brane, we could only find constant solutions of (\ref{susycondDp}). If we choose negative world volume chirality, they are $(+ \frac{1}{2}, - \frac{1}{2}, - \frac{1}{2},- \frac{1}{2}, + \frac{1}{2})$, $(+ \frac{1}{2}, - \frac{1}{2}, - \frac{1}{2},+ \frac{1}{2}, - \frac{1}{2})$, $(+ \frac{1}{2}, + \frac{1}{2}, + \frac{1}{2},+ \frac{1}{2}, - \frac{1}{2})$ and $(+ \frac{1}{2}, + \frac{1}{2}, + \frac{1}{2},- \frac{1}{2}, + \frac{1}{2})$ meaning that only 1/4 supersymmetries are preserved. Apparently for positive chirality the D5 brane breaks all the supersymmetries.

\textbf{D1}: here, one can check that no $u$-independent Killing spinors are allowed, independently of the world volume chirality (unless some $W_i$ is set equal to zero). The D1 brane seems not supersymmetric in our background.

In order to study the equations of motion we remark that $S$ appears only in
the $uu$ direction, so that for our ansatz the other equations are
automatically satisfied, therefore we will focus only on the equation of motion
for the $uu$ direction.

In order to analyse the equations of motion we found more convenient to work in the Einstein frame, where the metric is:

\begin{equation}
ds^{2}=H\left( \vec{y}\right) ^{\frac{p-7}{8}}\left( 2du\left( dv+S\left(
\vec{y}\right) du\right) -d\vec{x}^{\,2}\right) -H\left( \vec{y}\right) ^{\frac{p+1}{8}}d%
\vec{y}^{\,2}
\end{equation}
and the rest of the fields are identical to (\ref{stringf}).
In this frame the equations of motion are \cite{stelle}: 
\begin{equation}
\label{eom}
R_{\mu \nu }=\frac{1}{2}\partial _{\mu }\phi \partial _{\nu }\phi +S_{\mu
\nu}
\end{equation}
with 
\begin{multline}
S_{\mu \nu }=\sum_{p}\frac{1}{2\left( p+1\right) !}e^{\frac{\left(
3-p\right) \phi }{2}}\left( F_{\left[ p+2\right] \mu \alpha _{1}\dots \alpha
_{p+1}}F_{\left[ p+2\right] \nu }^{\alpha _{1}\dots \alpha _{p+1}} 
-\frac{p+1}{8\left( p+2\right) }g_{\mu \nu }F_{[p+2]}^2 \right)  
\end{multline}
with
\begin{equation}
F_{[p+2]}^2=F_{\left[ p+2\right]
\alpha _{1}\dots \alpha _{p+2}}F_{[p+2] }^{\alpha _{1}\dots \alpha _{p+2}}
\end{equation}
\qquad

The $uu$ component of the Ricci tensor for $p$ generic is:

\begin{equation}
R_{uu}=\frac{\partial _{a}\partial _{a}S}{H}-\frac{p-7}{8}S\frac{\partial
_{a}H\,\partial _{a}H}{H^{3}}+\frac{p-7}{8}S\frac{\partial _{a}\partial _{a}H%
}{H^{2}}
\end{equation}

From this we obtain the conditions:

\begin{subequations}
\begin{eqnarray}
\partial _{a}\partial _{a}H &=&0 \\
\partial _{a}\partial _{a}S &=&\frac{1}{2}%
(|W_{1}|^{2}+|W_{2}|^{2}+|W_{3}|^{2}+|W_{4}|^{2})
\end{eqnarray}
\end{subequations}

As before, the first condition says that $H$ is harmonic (in the transverse $%
\left( 9-p\right) $--dimensional space) while the second simply states a differential relation between $S$ and the $W^{\prime }$s. Taking into account the correct asymptotic behavior for $H$, one
must have:

\begin{equation}
H=1+\frac{Q_p}{r^{(7-p)}}
\end{equation}
for $p$ different from 7, and:

\begin{equation}
H=1+Q_7 \,\ln{r}
\end{equation}
for $p=7$, $r$ being the radius of the transverse space.

For the case of the D5-brane and for constant $W^{\prime }$s (hence $S$ has to be quadratic) our background is similar to those studied in \cite{BKP} in the context of $AdS_3 \times S^3 \times {\Bbb R}^4$ and its Penrose limits. Note, however, that we are considering more general backgrounds. 

We stress the fact that $S$ doesn't have to be function of all the transverse coordinates, that is, we don't have to give mass to all the transverse scalars. So for a Dp-brane solution with a given $S$, there will be also a Dp'-brane solution with $p'<p$.

%%%%%%%%%%%%%%%%%%%%%%%%%%%%%%%%%%%%%%%%%%%%%%%%%%%%%%%%%%%%%%%%%%%%%%%%%

\section{The D1/D5 system} 

%%%%%%%%%%%%%%%%%%%%%%%%%%%%%%%%%%%%%%%%%%%%%%%%%%%%%%%%%%%%%%%%%%%%%%%%

In this section we extend the previous analisys to the D1/D5 system. This system plays an important role in the AdS/CFT duality since its near horizon limit is of the form $AdS_3 \times S^3 \times M$ (see, for instance \cite{maldareview}\cite{GHN}); its Penrose limit was considered for instance in \cite{RT}\cite{KS}. See also \cite{boyda} \cite{herdeiro} .First we solve the equations of motion (in the Einstein frame) and after we briefly discuss the issue of supersymmetry. 

We propose the following ansatz:

\begin{subequations}
\label{d1d5}
\begin{eqnarray}
\label{metricd1d5}
ds^{2} &=& H_1^{-\frac{3}{4}} H_5^{-\frac{1}{4}}( 2du( dv+S(
\vec{y}) du)) -  H_1^{\frac{1}{4}}H_5^{-\frac{1}{4}}d\vec{x}^{\,2} -H_1^{\frac{1}{4}} H_5^{\frac{3}{4}}d\vec{y}^{\,2}\\
F_{[3]D} &=& du\wedge dv \wedge dH_1^{-1}
\\
F_{[7]D} &=& du\wedge dv\wedge dx^{1}\dots \wedge dx^{4}\wedge dH_5^{-1}
\\
F_{[5]B} &=&\frac{1}{\sqrt 2}(\varphi _{\lbrack 4]})\wedge du+c.c. \\
\varphi _{\lbrack 4]} &=&W_{1}d{\bar{z}}^{1}\wedge dz^{2}\wedge dz^{3}\wedge
dz^{4}+W_{2}dz^{1}\wedge {d\bar{z}}^{2}\wedge dz^{3}\wedge dz^{4}\\
e^{2\phi}&=&H_1 H_5^{-1}
\end{eqnarray}
\end{subequations}

Here $\vec{x}$ and $\vec{y}$ denote points of $\Bbb{R}^4$. $H_1$ and $H_5$ are only function of the coordinates $\vec{y}$ transverse to the D5-brane. As can be seen from the metric, we place the D1-brane on the $u,v$ coordinates and the D5-brane on $u,v$ and $\vec{x}$ so that the D1-brane lies inside the D5-brane.

As before, the only  non trivial equation of motion is the one in the $uu$ direction. From the metric (\ref{metricd1d5}) we have the following Ricci tensor:

\begin{equation}
R_{uu}=\frac{3}{4} \frac{S \partial_a H_1 \partial_a H_1}{H_5 H_1^3} - \frac{3}{4} \frac{S \partial_a \partial_a H_1}{H_5 H_1^2} +\frac{1}{4} \frac{S \partial_a H_5 \partial_a H_5}{H_1 H_5^3} -\frac{1}{4} \frac{S \partial_a \partial_a H_5}{H_5^2 H_1} + \frac{\partial_a \partial_a S}{H_1 H_5}
\end{equation} 

From the equations of motion (\ref{eom}) we obtain the following conditions:

\begin{subequations}
\begin{eqnarray}
\partial_a \partial_a H_1&=& 0\\
\partial_a \partial_a H_5 &=& 0\\
\partial_a \partial_a S &=& \frac{1}{2}(|W_1|^2 +|W_2|^2)
\end{eqnarray}
\end{subequations}

So, $H_1$ and $H_5$ must be harmonic in the transverse directions and the usual relation between $S$ and $W_i$ must be fulfilled.

One can check that the supersymmetry conditions reduce to (\ref{susycond}) provided $H$ is replaced by the product $H_1 H_5$, except in the $i$ direction (i.e. belonging to the D5-brane but not to the D1-brane), where the condition reads:

\begin{equation}
\partial_i \epsilon + H_5^{-1/2}W_5(i \sigma_2) \gamma_{\underline i} \epsilon = 0
\end{equation}

We stress that such equations were obtained supposing the spinor has definite chirality with respect to the worldvolume of both D-branes. As in the case of the D1-brane, apparently such conditions are too restrictive to allow any supersymmetry.

\section{Conclusions}

We have found a solution describing a fully localized D3-brane on a pp-wave
background. Such solution turns out to be supersymmetric, and it preserves,
apparently 1/8 supersymmetries. We also show how to generalize our
results to the case of other Dp-branes. Our conclusions are summarized in the table. Finally we have studied the D1/D5-brane system, which turns out to be not supersymmetric.

\begin{table}[th]
\begin{center}
\begin{tabular}{|c|c|c|c|c|}
\hline
  & D1 & D3 & D5 & D7 \\
\hline
$\Gamma_{wv}\epsilon=\epsilon$ & 0 susy  & 1/8 susy & 0 susy & 1/4 susy \\
\hline
$\Gamma_{wv}\epsilon=-\epsilon$ & 0 susy & 1/8 susy & 1/4 susy & 1/4 susy \\ 
\hline
\end{tabular}
\end{center}
\caption{Summary of the results for Dp-branes.}
\end{table}

As a possible further development, it could be interesting to study the
gauge theories living on the worldvolume of such D-branes, or
to study the backgrounds obtained from ours after some dualities. In this way other Dp and Dp/Dp' solutions are expected.

\noindent{\bf Acknowledgments} 

We thank Edi Gava and Kumar Narain for having guided us through this work. We also thank Carlo Maccaferri, Davide Mamone and especially Patrick Meessen for enjoyable discussions.

\end{document}